\documentclass[10pt]{iopart}
\usepackage{graphicx}
\begin{document}
\def\PRL{Phys. Rev. Lett. }
\def\PRC{Phys. Rev. C }
\def\PRD{Phys. Rev. D }
\def\PLB{Phys. Lett. B }
\def\NPA{Nucl. Phys. A }
\def\pT{\mbox{$p_T$}}
\def\pt{\mbox{$p_T$}}
\def\v2{\mbox{$v_2$}}
\def\sqrtsNN{\mbox{$\sqrt{s_{NN}}$}}
\def\sqrts{\mbox{$\sqrt{s}$}}
\def\jt{\mbox{$j_T$}}
\def\kt{\mbox{$k_T$}}
\def\et{\mbox{$E_T$}}
\def\mkt{\mbox{$\langle k_T\rangle$}}
\def\met{\mbox{$\langle E_T\rangle$}}
\def\mjt{\mbox{$\langle j_T\rangle$}}

\title[Dihadron correlations at high \pT]{Dihadron correlations at high \pT}

\author{Kirill Filimonov}

\address{Nuclear Science Division, Lawrence Berkeley National Laboratory, 1 Cyclotron Road, Berkeley CA 94720}

\ead{KVFilimonov@lbl.gov}

\begin{abstract}
Jet quenching in the matter created in high energy 
nucleus-nucleus collisions provides a tomographic tool to probe
the medium properties. 
Recent experimental results from the Relativistic Heavy-Ion Collider (RHIC)
on characterization of jet production via dihadron correlations 
at high transverse momentum are reviewed. Expectations from the dihadron
measurements for the lower energy \sqrtsNN=62.4 GeV RHIC run are discussed.
\end{abstract}



%\maketitle

\section{Introduction}
Jets of hadrons come from fragmentation of high energy quarks
or gluons scattered with a large momentum transfer.
Jets produced in ultra-relativistic heavy-ion collisions 
at RHIC probe nuclear matter at extreme conditions of 
high energy density. 
Energetic partons propagating through the medium are predicted
to lose energy via induced gluon radiation, with the energy loss
depending strongly on the color charge density of the created system and the
traversed path length \cite{quenching}.
Attenuation of jets enables tomographic analysis of the
created matter  using the jet  
effectively as an external probe of the medium \cite{tomography}. 
At RHIC, high transverse momentum (\pt) measurements of hadron production 
have revealed three different observations related to jet quenching: 
strong suppression of the inclusive hadron spectra \cite{suppression}, 
large azimuthal anisotropies with respect to the reaction plane 
orientation \cite{flow}, 
and strong suppression of the back-to-back azimuthal correlations \cite{btob}.
Recent high \pt~measurements performed with d+Au collisions \cite{dAuPhobos,dAuPhenix,dAuStar,dAuBrahms} confirm
that the nuclear attenuation observed in central Au+Au collisions is due to
final-state interactions of jets in the dense matter formed in heavy ion 
collisions. A comprehensive review of the RHIC data and theory is given in \cite{Jacobs:2004qv}, while most recent experimental results on jet production at RHIC
are summarized in \cite{Filimonov:2004qz}. I will review the 
dihadron correlation measurements at high \pT~at the top RHIC 
energy (\sqrtsNN=200 GeV) and discuss 
the expectations from the
lower energy \sqrtsNN=62.4 GeV run.

\section{Suppression of away-side correlations: jet quenching}

Hard scattered partons fragment
into a high energy cluster (jet) of hadrons which are 
distributed in a cone of size $\Delta\eta\Delta\phi=\sim0.7$ 
in pseudorapidity and azimuth.
The large multiplicities in nuclear collisions make full jet reconstruction
impractical. Instead, correlations of high \pT~hadrons
are used for the identification of jets on a statistical basis. 
The relative azimuthal angle distributions of dihadrons reveal jet-like correlations that are 
characterized by the peaks at $\Delta\phi=0$ 
(near-side correlations) 
and at $\Delta\phi=\pi$ (back-to-back).  

\begin{figure}[ht]
\begin{center}
\begin{flushleft}{\includegraphics*[%
  keepaspectratio,
  width=0.48\columnwidth]{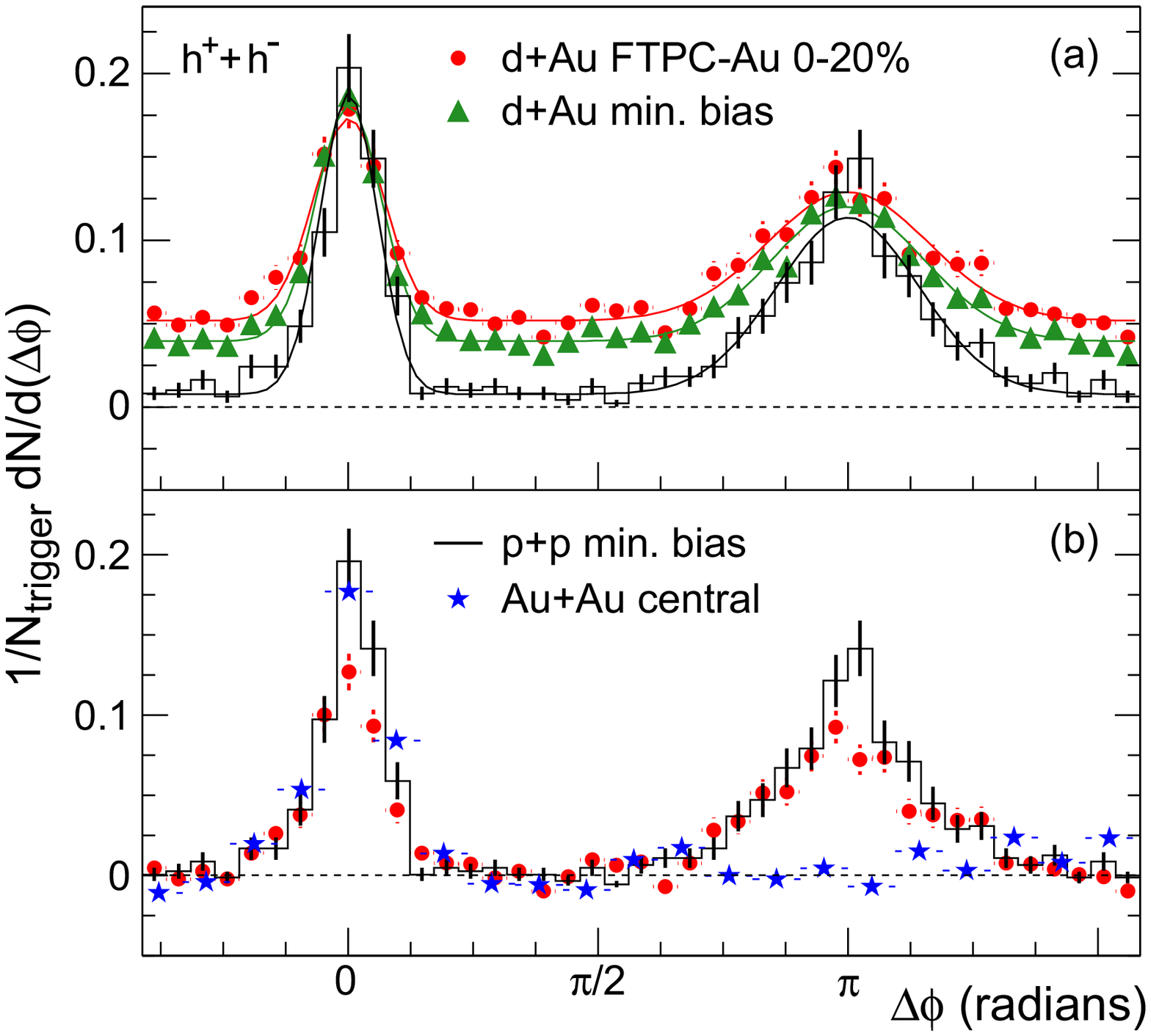}}\hfill{\includegraphics*[%
  width=0.48\columnwidth]{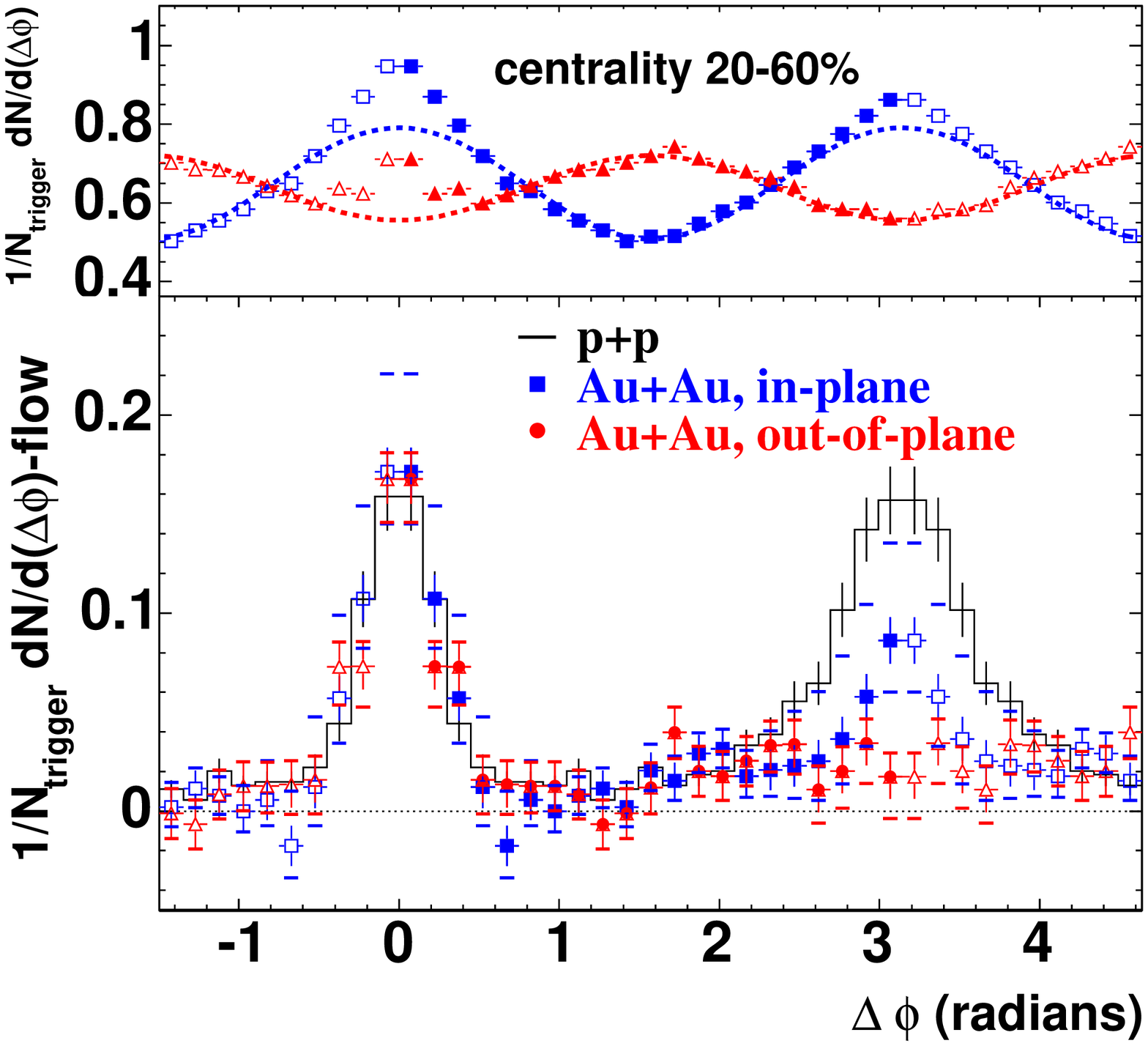}}\end{flushleft}
\caption{Left: STAR data \cite{dAuStar} on dihadron azimuthal 
correlations in p+p, d+Au and central Au+Au collisions. Right:
STAR data \cite{inoutStar} on modification of the dihadron correlations
in Au+Au collisions with respect to reaction plane, compared to p+p.
\label{fig:dAu}}
\end{center}
\vspace{-0.5cm}
\end{figure}

Striking evidence of in-medium effects on dihadron correlations is presented
in Figure~\ref{fig:dAu}. The left panel shows the relative azimuthal 
angular distribution between trigger hadrons of 
$p_T^{\rm trig}$=4-6 GeV/c and associated hadrons 
(2 GeV/c$<p_T^{\rm assoc}<p_T^{\rm trig}$)  
measured in central Au+Au collisions, compared to that 
in p+p and d+Au collisions.
Correlations of high \pT~hadrons at small relative angles are seen to be 
essentially 
unaffected by the medium (the strength of the near-side correlations 
is consistent with that measured in p+p and d+Au collisions).
In sharp contrast, the away-side (back-to-back) correlations are strongly suppressed in 
the most central Au+Au collisions. The right panel shows the variation
of back-to-back suppression with azimuthal orientation of 
the jets relative to the reaction plane for non-central Au+Au collisions.
Elliptic flow subtracted distributions \cite{inout} show larger
suppression of the back-to-back correlations for the 
out-of-plane trigger particles
than for in-plane.
%-----------------------------------------------------------------------
%-----------------------------------------------------------------------
These observations are naturally predicted by jet quenching 
models, where the energy  
loss of a parton depends on the density of and distance traveled 
through the medium.
The high \pt~trigger biases the initial production point to be near 
the surface so the near-side correlations should be similar to those
seen in p+p collisions. The away-side correlations are suppressed 
in the dense medium, and more suppressed 
when the trigger hadron is emitted perpendicular to the reaction plane.

\section{Enhancement of near-side correlations: recombination of shower partons with thermal quarks?}
\begin{figure}[ht]
\begin{center}
\begin{flushleft}{\includegraphics*[%
  keepaspectratio,
  width=0.51\columnwidth]{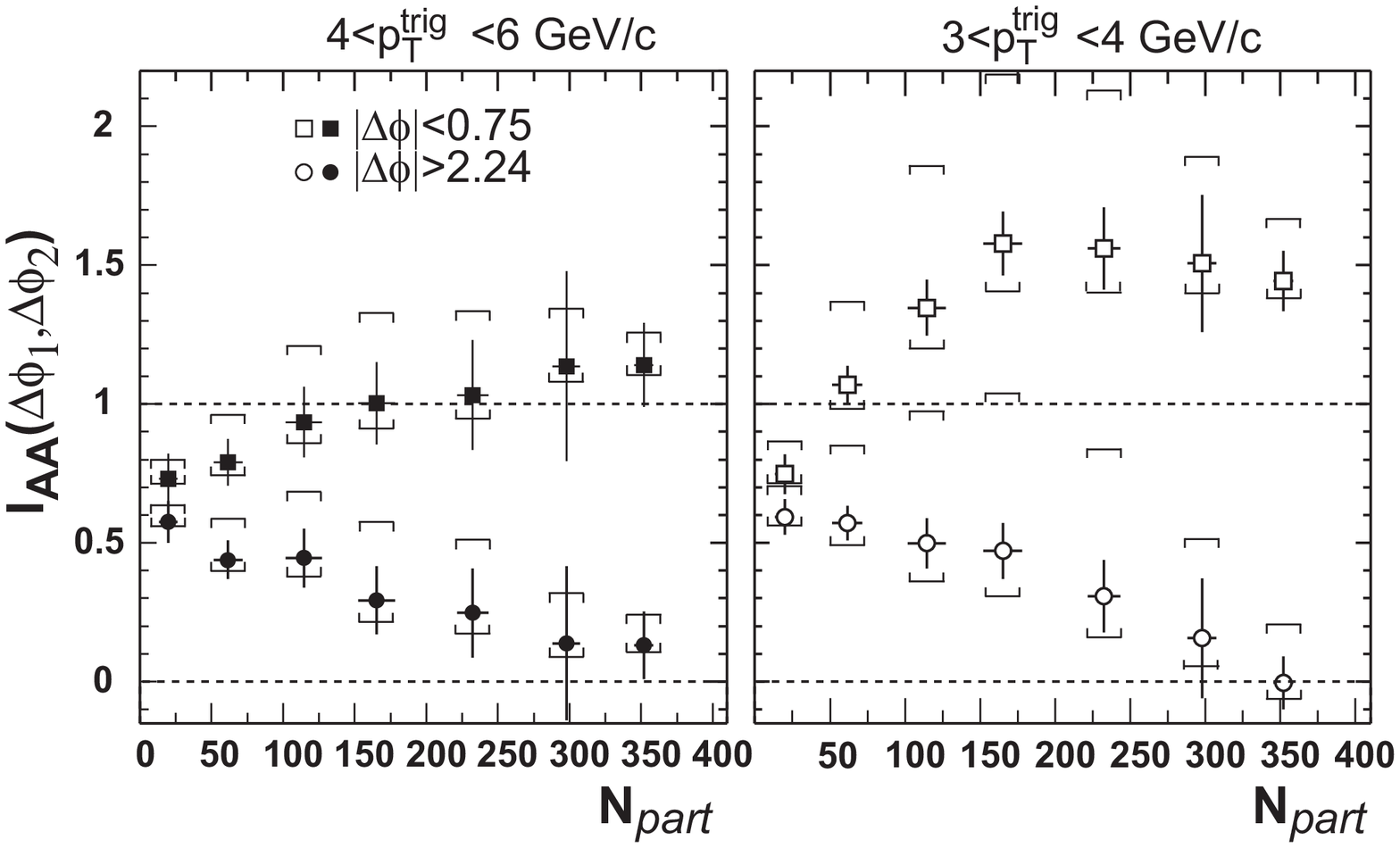}}\hfill{\includegraphics*[%
  height=2.0in,
  width=0.44\columnwidth]{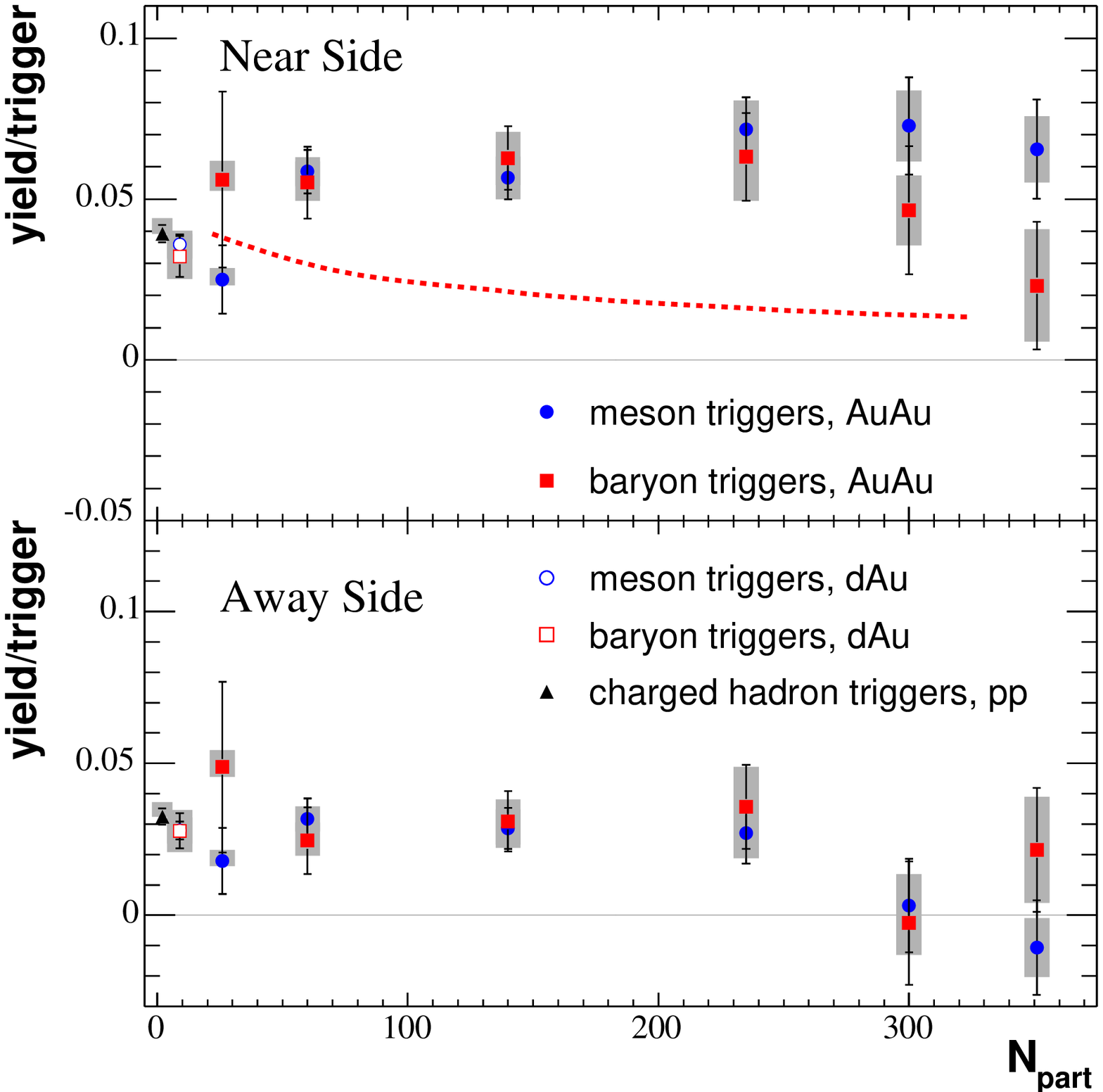}}\end{flushleft}
\caption{Left: STAR data \cite{btob} on AA/pp ratios of 
the near ($|\Delta \phi|<0.75$ rad) and away-side ($|\Delta \phi|>2.24$) 
correlations in Au+Au collisions. Right:
PHENIX data \cite{corrPhenix} on the associated yield per trigger 
for the near and away-side correlations in p+p, d+Au and Au+Au
collisions.  
\label{fig:iaa}}
\end{center}
\vspace{-0.5cm}
\end{figure}

One of the puzzles of the dihadron correlation measurements
is the observation of stronger (compared to p+p) 
correlations on the near-side for trigger and associated hadrons at lower
\pT. In the left panel of Figure~\ref{fig:iaa}, the ratio of associated yields 
%-----------------------------------------------------------------------
%-----------------------------------------------------------------------
measured in Au+Au and p+p collisions for trigger particle intervals 
$p_T^{\rm trig}$=4-6 GeV/c and $p_T^{\rm trig}$=3-4 GeV/c is shown.
The ratio should be unity
if the hard-scattering component of
Au+Au collisions is simply a superposition of p+p collisions unaffected
by the nuclear medium.  
The away-side correlations are suppressed in the most central collisions
for both intervals of $p_T^{\rm trig}$. However, for the lower
value of $p_T^{\rm trig}$, the
near-side correlations show an enhancement of about 50\% in 
mid-central Au+Au collisions
compared to p+p. This enhancement is also observed
for baryon and meson triggers with $p_T^{\rm trig}$=2.5-4.0 GeV/c which
are associated with the unidentified charged particles of $p_T^{\rm assoc}$=1.7-2.5 GeV/c 
(right panel of Figure~\ref{fig:iaa}). The near-side associated yields
per baryon/meson trigger almost double in mid-central Au+Au collisions
compared to p+p and d+Au. These observations are incompatible with
vacuum fragmentation of hard scattered partons and do not emerge naturally  
from a model focusing on the medium modification
of the fragmentation function due to energy loss.
At the same time, other 
aspects of particle production in central Au+Au collisions at RHIC
in the \pt-range of 2-5 GeV/c are also incompatible with
jet fragmentation in simpler systems, such as large $p/\pi$ ratio 
 and different suppression of proton/pion \cite{baryonmeson} 
and lambda/kaon yields \cite{idflowstar}.
%and deviation from hydrodynamic mass-ordering in strength
%of elliptic flow of pions/protons \cite{idflowphenix} and 
%kaons/lambdas \cite{idflowstar}.
Models based on coalescence/recombination of thermal/shower partons 
\cite{Greco:2003mm,Fries:2003vb,Hwa:2004vi} are 
successful in qualitatively
describing these features \cite{fries}.
However, the models which assume 
no correlations among the quarks before recombination are
incompatible with dihadron measurements.
Currently, there are different theoretical approaches 
to reconcile the recombination mechanism of hadronization 
with experimental observations of finite dihadron correlations.
One approach is to treat the fragmentation process as the
result of recombination of shower partons created 
by a hard parton. In a heavy-ion collision environment, recombination 
of thermal and shower partons leads to the different structure
of jets from that produced in p+p collisions \cite{Hwa:2004sw}.
Such thermal-shower recombination is then expected to dominate
the shower-shower recombination (the only source of 
dihadron correlations in p+p collisions) in the intermediate
\pT~range.
Another recent suggested scenario is that correlated emission of hadrons
may arise from recombination of correlated partons from a quasi-thermal 
medium \cite{Fries:2004hd}. The model lacks 
quantitative description of the origin of correlations in the parton phase, but it does show that if correlations among quarks exist, they are amplified by the 
hadronization process via recombination.
These theoretical efforts may provide a plausible explanation 
for the dihadron correlation measurements in the intermediate
\pT~range. 

\section{What to expect from dihadron measurements at lower energies?}

\begin{figure}[ht]
\begin{center}
\begin{flushleft}{\includegraphics*[%
  keepaspectratio,
  width=0.48\columnwidth]{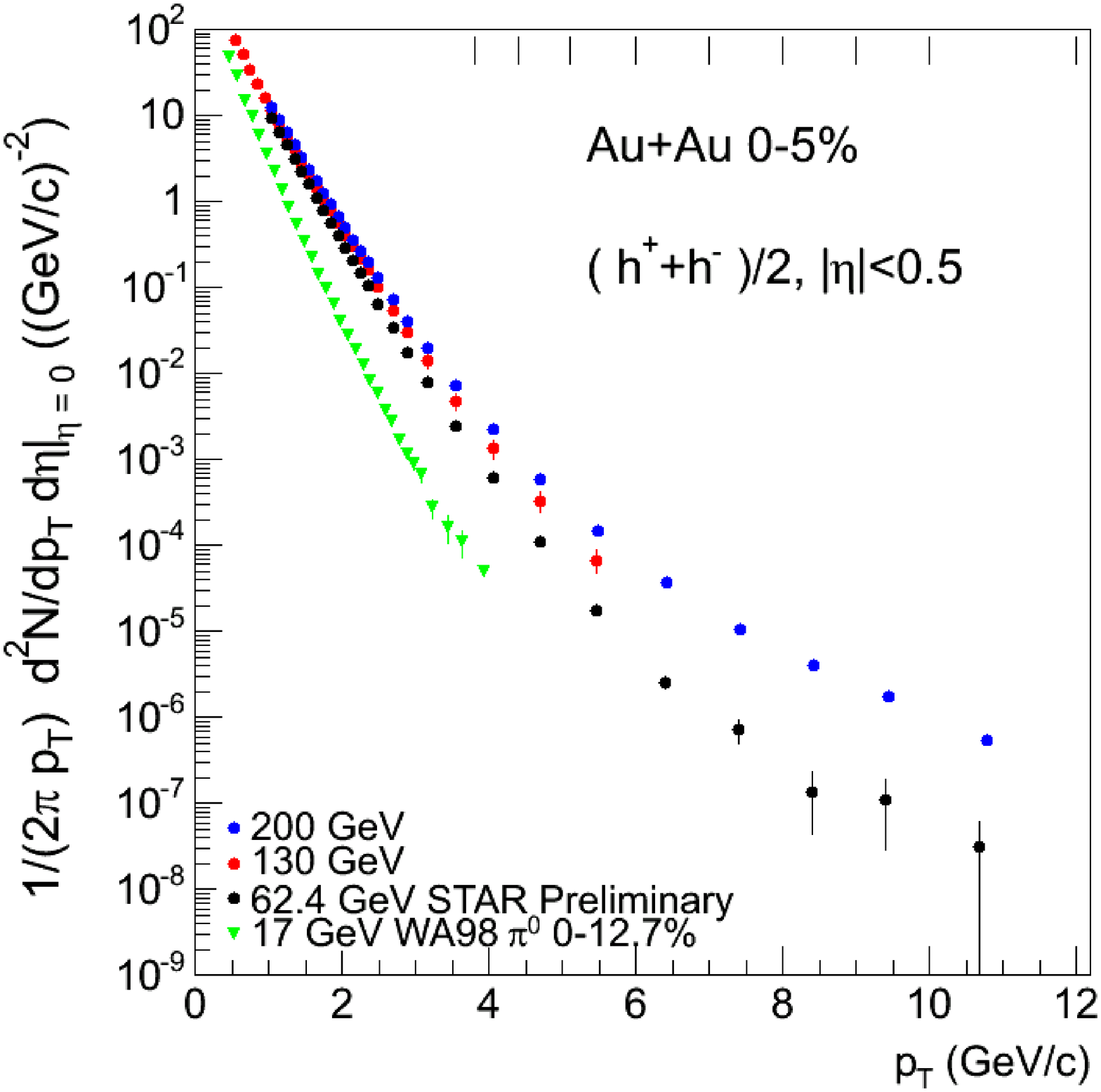}}\hfill{\includegraphics*[%
  height=2.1in,
  width=0.48\columnwidth]{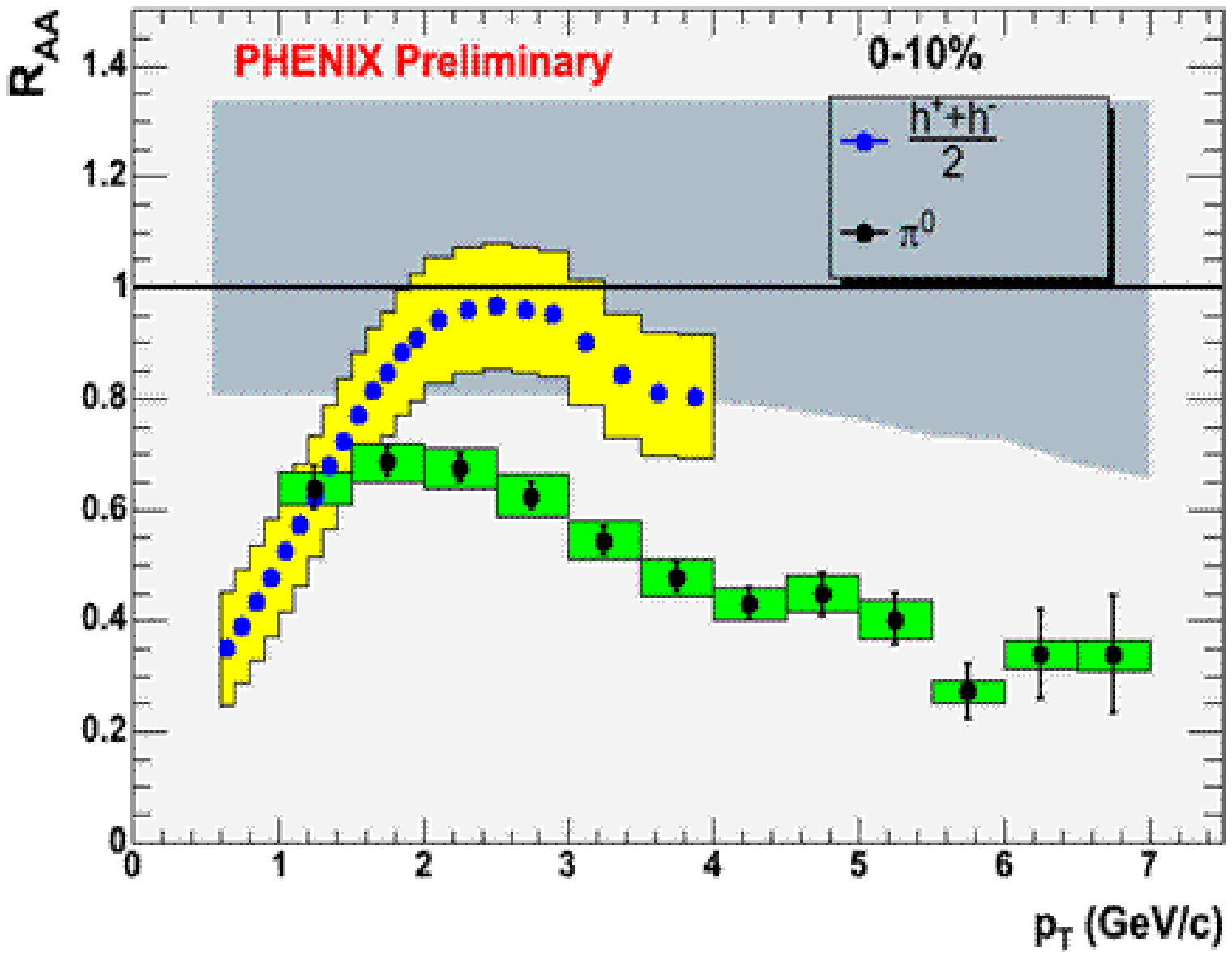}}\end{flushleft}
\caption{Left: STAR preliminary data \cite{STAR62} on the
inclusive charged hadron \pT~spectra in central Au+Au collisions at
 \sqrtsNN =62.4 GeV, compared to that measured at \sqrtsNN = 17, 130 and 200
GeV. Right:
PHENIX preliminary data \cite{Phenix62} on the nuclear modification factor
$R_{AA}$ measured for charged hadrons and 
$\pi^0$'s in central Au+Au collisions at
 \sqrtsNN =62.4 GeV. 
\label{fig:spectra}}
\end{center}
\vspace{-0.5cm}
\end{figure}

The yield of single 
inclusive high \pT\ hadrons has been observed to be substantially 
suppressed (by a factor of 3-5 for \pT$>$ 6 GeV/c) in the most central 
Au+Au collisions 
at \sqrtsNN = 130 and 200 GeV compared to the expectation from 
p+p collisions at the same beam energy.
Figure~\ref{fig:spectra} (left) shows the STAR preliminary inclusive
charged hadron spectra measured at  \sqrtsNN =62.4 GeV, 
compared to that measured at \sqrtsNN = 17, 130 and 200 GeV. 
%-----------------------------------------------------------------------
%-----------------------------------------------------------------------
Steeper spectra are measured for lower energies, similar to p+p collisions.
The right panel of Figure~\ref{fig:spectra} shows the PHENIX preliminary 
data on the nuclear modification factor ($R_{AA}$), or the ratio of 
nuclear geometry scaled yields 
measured in central Au+Au collisions to the 
p+p reference data. The observed suppression of the $\pi^0$ yield  
at  \sqrtsNN =200 GeV is similar to that measured
at the lower energy for $p_T>$5-6 GeV/c, as predicted \cite{quenching62}.

\begin{figure}[ht]
\begin{center}
\begin{flushleft}{\includegraphics*[%
  keepaspectratio,
  width=0.5\columnwidth]{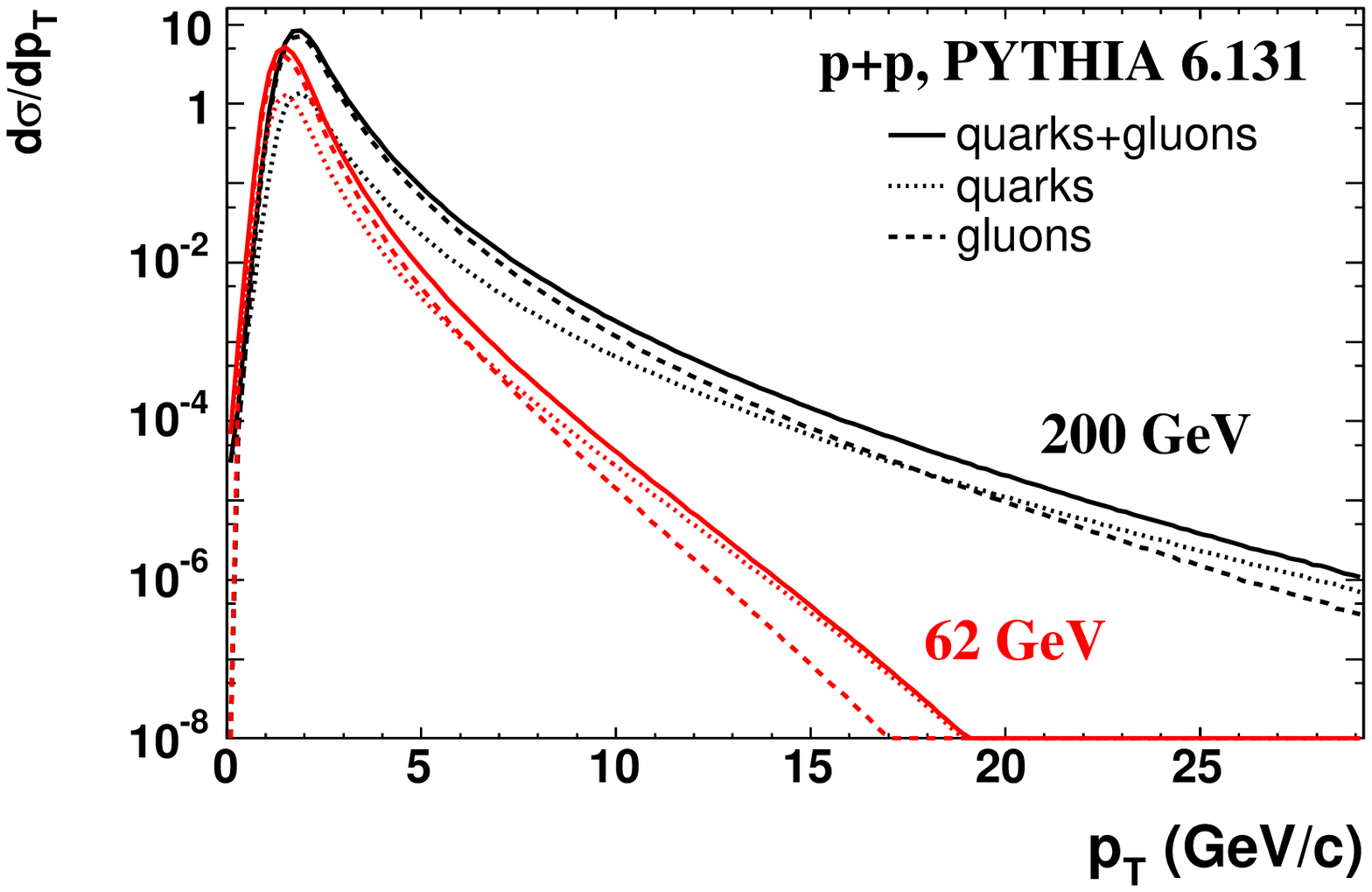}}\hfill{\includegraphics*[%
  width=0.5\columnwidth]{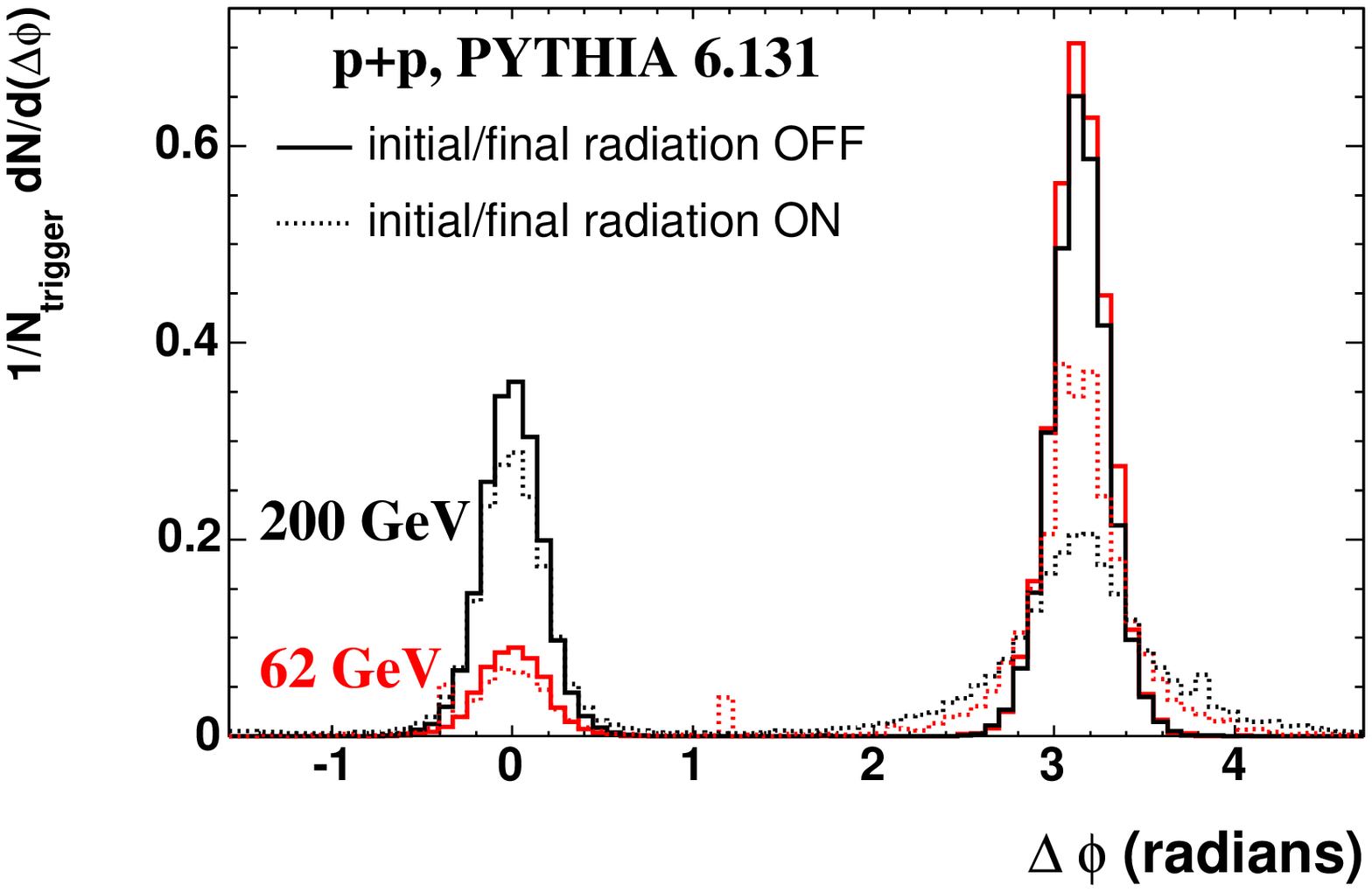}}\end{flushleft}
\caption{Left: Parton cross section calculated in PYTHIA for p+p collisions 
at \sqrts =62.4 and 200 GeV. Multiple interactions and initial and final state QCD radiation were turned off.  Right: Dihadron azimuthal distributions 
from PYTHIA for $p_T^{\rm trig}$=4-6 GeV/c and  2 GeV/c $<p_T^{\rm assoc}<p_T^{\rm trig}$.
\label{fig:pythia1}}
\end{center}
\vspace{-0.5cm}
\end{figure}

Let us examine whether steeper spectra measured at the lower energy
affect dihadron correlations. For this purpose we employed the
PYTHIA event generator (v. 6.131) \cite{pythia} for p+p collisions 
at \sqrts =62.4 and 200 GeV. The event generator was run with default 
parameters (only multiple interactions were switched off) and with and
without initial and final state QCD radiation (parton showers).
Figure~\ref{fig:pythia1} (left) shows the parton yields generated
by PYTHIA at \sqrts =62.4 and 200 GeV. Again, calculations show 
steeper parton spectra at the lower energy. For transverse momenta
about 10 GeV/c, quark scatterings dominate at \sqrts =62.4, whereas at
\sqrts =200 GeV contributions of quarks and gluons are comparable.

For dihadron correlations,
we selected trigger particles (charged pions, kaons, (anti)protons) with
$p_T^{\rm trig}$=4-6 GeV/c, and paired them with associated particles 
satisfying 2 GeV/c $<p_T^{\rm assoc}<p_T^{\rm trig}$. The particles were
restricted to $|\eta|<$0.9. 
Figure~\ref{fig:pythia1} (right) shows the azimuthal
distributions of associated particles in p+p collisions at  
\sqrts =62.4 and 200 GeV from PYTHIA.
The strength of the near-side correlations 
is much smaller (by about a factor of ~3) at \sqrts =62.4 GeV compared to
that at \sqrts =200 GeV, while the away side is similar for both energies. 
Initial and final state QCD radiation effects 
make correlations wider on the near and away sides for both energies and
also weaker on the away side. The reduction of the near-side 
correlation strength with 
energy, however, does not strongly depend on whether partons are allowed to 
shower in the calculation.
%-----------------------------------------------------------------------
%-----------------------------------------------------------------------

\begin{figure}[ht]
\begin{center}
\begin{flushleft}{\includegraphics*[%
  keepaspectratio,
  width=0.5\columnwidth]{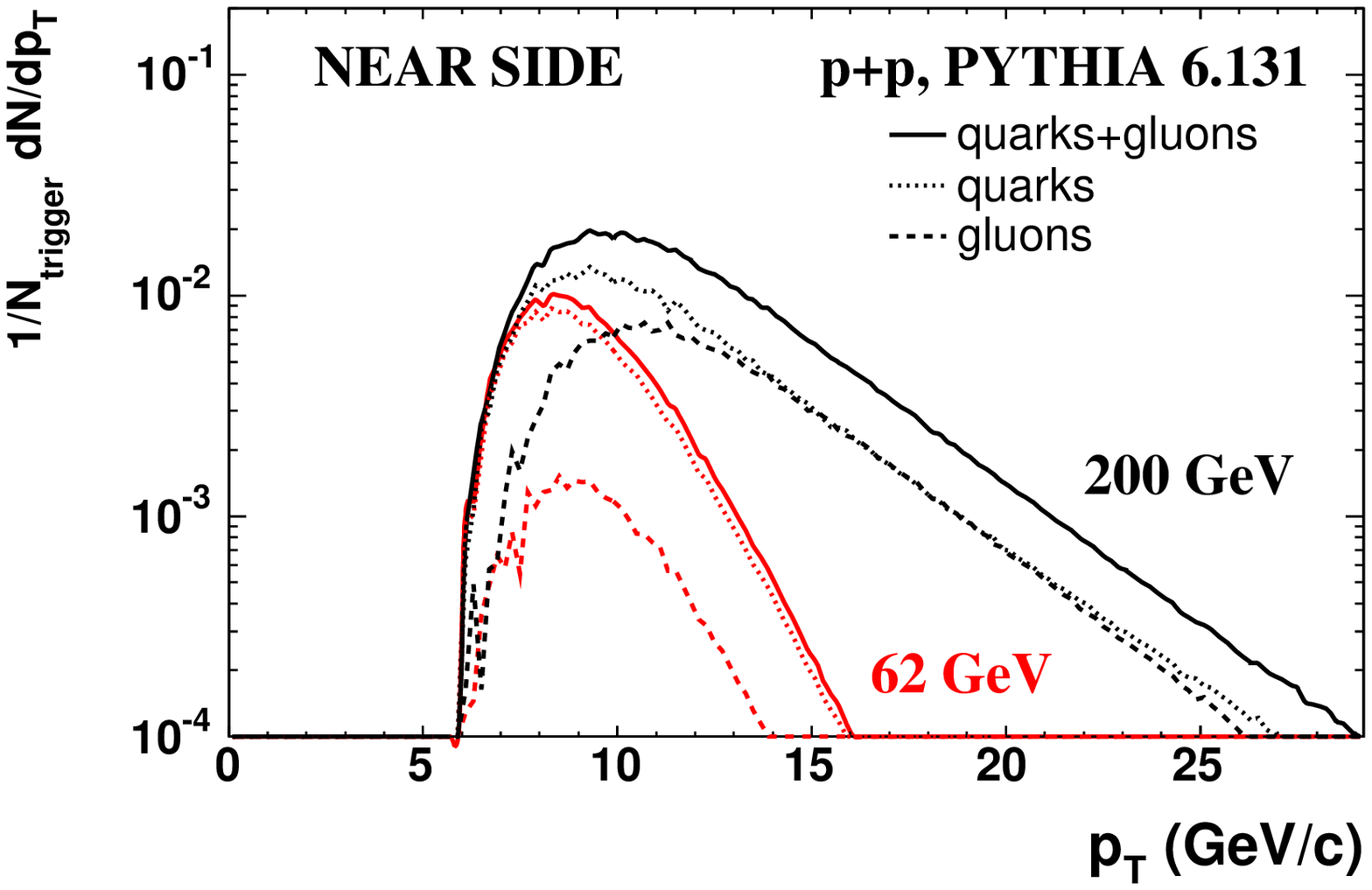}}\hfill{\includegraphics*[%
  width=0.5\columnwidth]{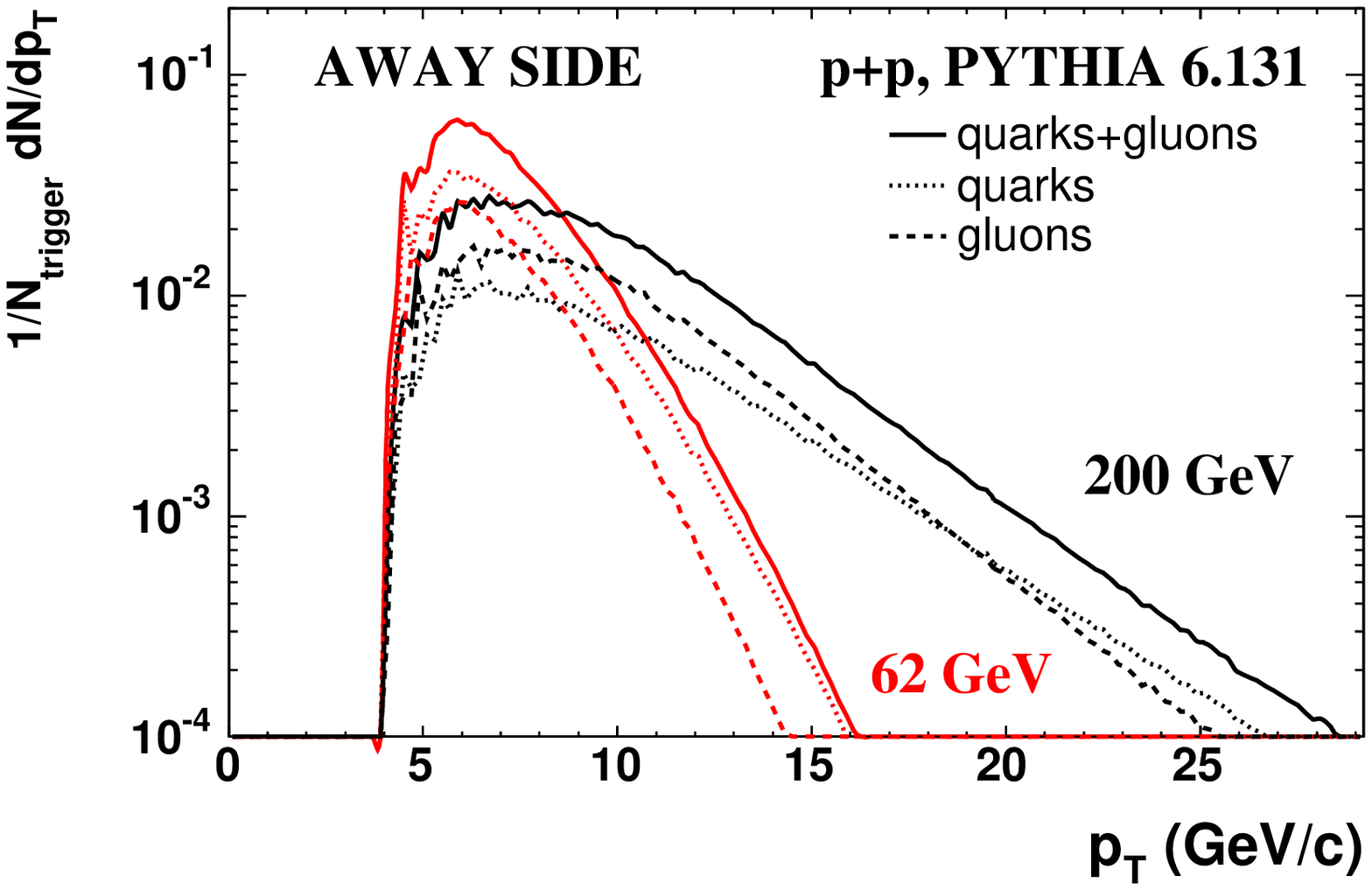}}\end{flushleft}
\caption{Parton \pT~ distributions associated with the near (left) and
away (right) side dihadron correlations.
\label{fig:pythia2}}
\end{center}
\vspace{-0.5cm}
\end{figure}

To gain a better understanding of this observation, 
we switched off the initial and final state QCD radiation in the generator 
(which makes unambiguous association of charged hadrons to their original parent parton
difficult) and studied the \pT~distributions of partons associated 
with the
near and away-side peaks separately (Figure~\ref{fig:pythia2}).
The distributions corresponding to the cases when trigger hadron originates
from the fragmentation of a quark or a gluon are also shown in 
Figure~\ref{fig:pythia2}. On the near side, the mean transverse momentum
of the parent parton is $\langle p_T \rangle$=11.7 GeV/c for 
\sqrts =200 GeV and 9.1 GeV/c at \sqrts =62.4 GeV.
On the away side, the mean transverse momentum
of the parent parton is $\langle p_T \rangle$=9.5 GeV/c for 
\sqrts =200 GeV and 7.0 GeV/c at \sqrts =62.4 GeV.
The difference of $\sim$2 GeV/c 
between the near and away side $\langle p_T \rangle$-values 
is due to the trigger bias in the dihadron correlations: 
to observe a near-side peak one needs a parton of at least \pT=6 GeV/c, 
fragmenting into the trigger hadron of \pT=4 GeV/c
and associated hadron of \pT=2 GeV/c, whereas on the away side the 
parton~\pT~can be just 4 GeV/c.
In addition, due to the different slopes of the parton \pT~distributions
at the two energies, both near and away side correlations for fixed $p_T^{\rm trig}$ and $p_T^{\rm assoc}$ correspond to larger (by about $\sim$ 2 GeV/c) original parton \pT~at higher energy. From the parton distributions shown in 
Figure~\ref{fig:pythia1} one can show that the ratio of the cross sections
corresponding to the near and away $\langle p_T \rangle$'s is about 0.3 at 
\sqrts =200 GeV and about 0.1 at \sqrts =62.4 GeV, or a factor of $\sim$3
between the two energies. 
The difference in the steepness of parton cross sections at the two energies
causes the change in the fragmentation of those partons contributing trigger 
particles: for fixed $p_T^{\rm trig}$=4-6 GeV/c,  
the fraction of the momentum of the parent parton
carried by the trigger hadron is  $\langle z_{\rm trig} \rangle$=0.64 at 200 GeV and 0.76 at 62 GeV.
%-----------------------------------------------------------------------
%-----------------------------------------------------------------------
Higher $\langle z \rangle$ of trigger particles at lower energy reduces the probability that an associated particle will
satisfy the threshold $p_T^{\rm assoc}$ concurrently.

\begin{figure}[b]
\begin{center}
\includegraphics[
  keepaspectratio,
  width=0.8\columnwidth]{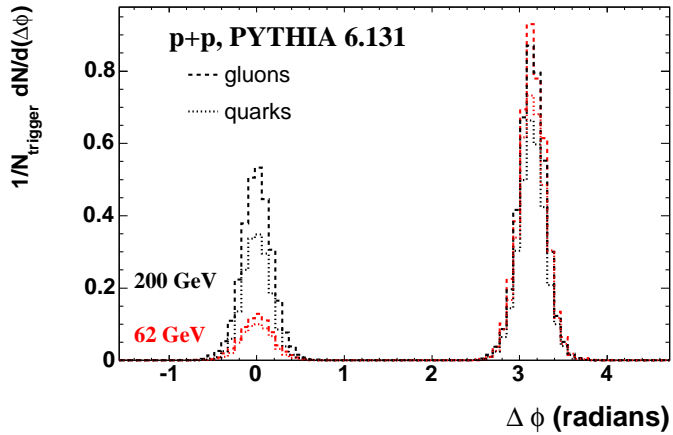}
\caption{Dihadron azimuthal distributions 
from PYTHIA for trigger hadrons originating from quark or gluon fragmentation.
\label{fig:pythia3}}
\end{center}
\vspace{-0.5cm}
\end{figure}

Another potentially important difference between the two energies is the lower value of the Bjorken x=0.1
at higher energy (compared to x=0.2-0.3 at 62.4 GeV) 
for $p_T^{\rm trig}$=4-6 GeV/c and $p_T^{\rm assoc}$=2-4 GeV/c. 
One can see from Figure~\ref{fig:pythia2} that the near-side correlations
of  hadrons 
at \sqrts =62.4 GeV come predominantly from fragmentation of quarks, while
at \sqrts =200 GeV there is a sizable contribution from gluons (about 40\%).
Gluon fragmentation is characterized by larger hadron multiplicities
which should appear as stronger near-side correlations in the dihadron 
distributions. Figure~\ref{fig:pythia3} compares 
the dihadron azimuthal distributions 
from PYTHIA for trigger hadrons originating from either 
quark or gluon fragmentation.
%-----------------------------------------------------------------------
%-----------------------------------------------------------------------
Indeed, stronger correlations are observed for gluon fragmentation, but
the gluon/quark ratios at the two energies cannot fully account for
the ``suppression'' of the near side correlations at \sqrts =62.4 GeV.
Thus we conclude that the predominant reason for the weaker near-side
correlations at the lower energy is the steepness of the parton cross section.
This effect should also exhibit itself for the dihadron measurements in
Au+Au collisions. The fact that at lower energy 4 GeV/c hadrons 
come primarily from
fragmentation of quarks, while at higher energy they come from an approximately
equal mix of quarks and gluons, may help in understanding 
the mechanism of the partonic
energy loss in heavy-ion collisions.

\section{Acknowledgments}
We thank Abhijit Majumder, Urs Wiedemann and Ivan Vitev
for valuable discussions. 
This work was supported by the Director, Office of Science, Nuclear Physics, 
U.S. Department of Energy under Contract DE-AC03-76SF00098.

\end{document}